\documentclass[]{aa} 
\usepackage{graphicx}
\usepackage{natbib}  
\bibpunct{(}{)}{;}{a}{}{,}
\usepackage{txfonts}
\usepackage{hyperref}
\hypersetup{bookmarks=true,colorlinks=true,linkcolor=black,citecolor=blue,urlcolor=blue}

\begin{document}

 \title{Gaia and VLT astrometry of faint stars: Precision of Gaia DR1 positions and updated VLT parallaxes of ultracool dwarfs
  \thanks{\textit{Based on observations made with ESO telescopes at the La Silla Paranal Observatory under programme IDs 086.C-0680, 087.C-0567, 088.C-0679, 089.C-0397, and 090.C-0786.}}}
\titlerunning{Gaia and VLT astrometry of faint stars}

\author{P. F. Lazorenko\inst{1}
		\and J. Sahlmann\inst{2}}

\institute{Main Astronomical Observatory, National Academy of Sciences of the Ukraine, Zabolotnogo 27, 03680 Kyiv, Ukraine\\
		\email{laz@mao.kiev.ua}				
		\and
		Space Telescope Science Institute, 3700 San Martin Drive, Baltimore, MD 21218, USA}

\date{Received ; accepted  }
\abstract{}
{We compared positions of the Gaia first data release (DR1) secondary data set at its faint limit with CCD positions of stars in 20 fields  observed with the VLT/FORS2 camera. {The FORS2 position uncertainties are smaller than one milli-arcsecond (mas) and allowed us to perform an independent verification of the DR1 astrometric precision.}}
{In the fields that we observed with FORS2, we projected the Gaia DR1 positions into the CCD plane, performed a polynomial fit between the two sets of matching stars, and carried out statistical analyses of the residuals in positions.}
{The residual RMS roughly matches the expectations given by the Gaia DR1 uncertainties, where we identified three regimes in terms of Gaia DR1 precision: for $G\simeq17-20$ stars we found that the formal DR1 position uncertainties of stars with DR1 precisions in the range of 0.5--5\,mas are underestimated by $63  \pm  5$\%, whereas the DR1 uncertainties of stars in the range $7-10$\,mas are overestimated by a factor of two. For the best-measured and generally brighter $G \simeq 16-18$\, stars with DR1 positional uncertainties of $<0.5$\,mas, we detected $0.44 \pm 0.13$\,mas excess noise in the residual RMS, whose {origin can} be in both FORS2 and Gaia DR1. By adopting Gaia DR1 as the absolute reference frame we refined the pixel scale determination of FORS2, { leading to minor updates to the} parallaxes of 20 ultracool dwarfs that we published previously. We also updated {the FORS2} absolute parallax of the Luhman 16 binary brown dwarf system to $501.42\pm 0.11$\,mas.}
{}

\keywords{Astrometry  --  methods: data analysis -- surveys -- catalogs--parallaxes--stars: individual: WISE J104915.57-531906.1}
\maketitle

\section{Introduction}{\label{notes}}
Gaia's first data release (DR1) provides accurate astrometric and photometric data for about one billion stars in the magnitude range $G\simeq3-20.7$ \citep{DR1_coll,GaiaCollaboration:2016aa}. Independently of the mission's own validation effort \citep{gaia_validation}, several studies found generally excellent agreement between external measurements and  parallaxes \citep[e.g.][]{Gaia-Collaboration:2017aa,Casertano:2017aa} and proper motions \citep[e.g.][]{van-der-Marel:2016aa, Watkins:2017aa} of bright stars in the Gaia DR1 primary data set \citep{DR1}. The bulk content of DR1, however, is the secondary data set of generally fainter stars for which it lists stellar positions at the epoch 2015, but no parallax or proper motions. 

The {DR1} positions were  tested with numerous methods, including the comparison with external catalogs of sufficient precision \citep{gaia_validation}. The external validation of astrometric accuracy of positions was made with two catalogs: the URAT1 catalog \citep{URAT1} with a  precision of positions 10--30 milli-arcsecond (mas),  and with positions of quasars in the catalog  ICRF2 QSO \citep{fey}, and no deviations from the model of uncertainties adopted in DR1 were found.

\citet{Mignard2016} compared ICRF2 positions \citep{ma2009, fey} of  $G=16-20$ sources with the auxiliary quasar table of DR1. They reported that if the comparison is made with the secondary data set of DR1, the dispersion of  the normalised coordinate differences is about  30\% higher than expected for the defining ICRF2 sources,  probably because the positional uncertainties in  DR1 are underestimated.

So far, no other validation of the DR1 secondary data set astrometry was made at the mas-level because most external catalogs have insufficient accuracy.  Here, we address this issue using CCD astrometric data sets obtained with the FORS2 camera installed at the Very Large Telescope (VLT) as described by \cite{PALTA2}. Our ground-based astrometry has a typical precision of 0.1--0.5~mas for individual stars. Comparison with DR1 is made by projecting its star positions into CCD space, computing the residuals of positions, and performing the statistical analysis of the scatter of these residuals. This investigation concerns only the random component of astrometric errors, because FORS2 astrometry is inherently differential, and is applied to  $G\simeq16-20$ stars that are towards the faint end of the DR1 secondary data set.  

We also better characterise the distortion of the FORS2 camera which allows us to re-calibrate the pixel scale and to update the parallaxes of ultracool dwarfs observed {in our programs}.

\section{ Differential CCD data sets  of  FORS2 field stars}{\label{cat_raw}}
In 2010 we started an astrometric planet search targeting 20 ultracool  dwarfs  and very low-mass stars \citep{Palta1, sahlmann2015DE0823, Sahlmann_palta3}. Observations were obtained with the FORS2/VLT camera \citep{FORS}, whose focal plane is composed of two  CCD chips. { Images obtained with the high-resolution collimator have an approximate pixel scale of $s_0=0.1261\arcsec$/px \citep{PALTA2}}. Our observations design and reduction methods reach precisions of $\sim$0.05--0.07~mas for $I=15-17$~mag stars \citep{Lazorenko2009, PALTA2}. Our CCD data set { based on observations in 2010--2013} contains images of  $I\simeq16-22$~mag stars in 20 fields close to the southern galactic plane, each covering $\sim 4\arcmin\times4\arcmin$. The list of fields { with a numbering adopted in \citep{PALTA2} is given in Table~\ref{table}}.

\begin{table}[tbh]
\caption{Sky fields, RMS of the positional differences  between DR1 and  FORS2 data sets, and average values of $\sigma_\mathrm{ G}$ and $\sigma_\mathrm{ G}$ for  chip1. 
}
{\tiny
\centering
\begin{tabular}{@{}r|c@{\quad}c@{\quad}c@{\quad}c@{\quad}c|   c@{\quad}c@{\quad}c@{\quad}c@{\quad}c@{}}
\hline
\hline
  &   \multicolumn{5}{c|}{ comparison between DR1 and FOV  }&  \multicolumn{5}{c}{ comparison between DR1 and RF  }  \rule{0pt}{11pt}\\  
\cline{2-11}
Nr &  $N$      &$ n$ &  RMS    &$\langle \sigma_\mathrm{ G}\rangle$   &$\langle \sigma_\mathrm{ F}\rangle$  & $N$   &$n$  &    RMS     &$\langle \sigma_\mathrm{ G} \rangle$  & $\langle \sigma_\mathrm{ F}\rangle$     \rule{0pt}{11pt}   \\
   &           &     & (mas)   & (mas)               & (mas)           &       &     &   (mas)    &     (mas)          & (mas)                  \\
\hline                                                                                                                                                                                      
 1&       41   & 4   &  3.78   &      2.50            &    0.56        &   43    & 4   &    2.72   &       2.45         &    0.51               \rule{0pt}{11pt}\\
 2&      35   & 3   &  4.69   &      5.44            &    0.34        &   31    & 4   &    3.93   &       5.72         &    0.31               \\
 3&      25   & 3   &  2.80   &      4.25            &    0.70        &   19    & 3   &    3.98   &       4.86         &    0.70               \\
 4&      30   & 3   &  6.22   &      3.73            &    0.78        &   31    & 4   &    1.89   &       1.35         &    0.60               \\
 5&      55   & 4   &  4.05   &      3.64            &    0.72        &   53    & 4   &    3.61   &       3.67         &    0.67               \\
 6&      79   & 6   &  5.43   &      4.61            &    0.83        &   50    & 4   &    4.28   &       4.74         &    0.80               \\
 7&      97   & 6   &  4.16   &      2.91            &    0.66        &   33    & 4   &    2.09   &       3.52         &    0.63               \\
 8&      74   & 6   &  3.35   &      3.94            &    0.58        &   59    & 4   &    2.14   &       4.23         &    0.54               \\
 9&      51   & 5   &  2.83   &      2.23            &    0.32        &   50    & 5   &    2.21   &       2.24         &    0.30               \\
10&      29   & 3   &  5.48   &      5.49            &    0.53        &   27    & 3   &    4.35   &       5.62         &    0.50              \\
11&     130   & 5   &  3.54   &      4.06            &    0.69        &   65    & 4   &    1.83   &       3.61         &    0.67               \\
12&      51   & 4   &  2.95   &      4.02            &    0.48        &   39    & 4   &    2.20   &       3.43         &    0.46               \\
13&     116   & 7   &  4.23   &      3.88            &    1.03        &   67    & 4   &    2.64   &       3.61         &    0.75               \\
14&     168   & 6   &  6.17   &      4.68            &    0.76        &   16    & 3   &    5.20   &       4.99         &    0.84               \\
15&     122   & 6   &  4.37   &      2.90            &    0.67        &   38    & 5   &    3.20   &       3.13         &    0.67               \\
16&     285   & 7   &  7.79   &      5.54            &    0.88        &   71    & 3   &    7.87   &       5.01         &    0.81               \\
17&     274   & 7   &  5.96   &      3.98            &    0.56        &   49    & 3   &    5.41   &       3.85         &    0.52               \\
18&     261   & 7   &  5.81   &      4.70            &    0.57        &   71    & 3   &    4.34   &       4.56         &    0.54               \\
19&     178   & 7   & 10.88   &      9.16            &    1.11        &   42    & 4   &    2.81   &       6.67         &    0.92               \\
20&     115   & 5   &  3.92   &      9.84            &    0.48        &   74    & 3   &    4.02   &      10.21         &    0.47               \\
\hline                 
\end{tabular}   
\tablefoot{ $N$ is the number of common stars;  $n$ is the power of the fit 
model
} \\       
\label{table}          
}                      
\end{table}

The main application of this data set is the search for { reflex motion of the targets measured relative to surrounding stars in the plane of the sky.} To ensure the best elimination of the atmospheric image motion and optical distortion of the telescope,  the positions of the targets were computed relative to  a reference frame formed by the dense grid of reference stars.  The reference areas are circular and centred on the target. The reduction within each { individual CCD image} was made using a polynomial model with $  M = (n+1)(n+2)/2$ basic functions { per each CCD axis} which include full two-dimensional polynomials of $x$, $y$ with the maximum power $n$,  removing  in this way the most important low-frequency components of the image motion spectrum \citep{Lazorenko2009}. Besides the reduction within the CCD plane, this model also takes into account the change of positions due to the proper motion, parallax, and colour effects. 

\subsection{ RF data set  of  FORS2 field stars}{\label{cat_raw_rf}}
A first supplementary result of the program are the astrometric parameters of every star measured relative to the reference frame centred on the target ultracool dwarf: CCD positions $x$, $y$  { given in epoch 2011.38}, proper motions in the CCD system, parallaxes, and chromatic parameters. The precision of our stellar positions  $\sigma_\mathrm{ F} \sim0.1-1$\,mas is comparable to that of Gaia DR1 secondary data set, which is $0.1-20$\,mas  \citep{DR1_coll}, so we use the results of the $n=4$ ($M=15$) reduction for the current investigation. We will refer to it as the reference frame (RF) data set and it contains 6208 stars in 20 fields. There are two sub-sets, one per CCD chip.

The argumentation in the next sections relies on the fact that the system of these astrometric parameters   is strictly homogeneous because it is defined by the same {\it reference frame and  basic functions.} It means that for {each RF star} the differential $x$, $y$ and the 'absolute' $x_\mathrm{ abs}$, $y_\mathrm{ abs}$ positions in some external catalog given in  ICRF system, e.g.\ Gaia, are related by the expression 
\begin{equation}
\label{eq:abs}
x_\mathrm{F}  -   x_\mathrm{ abs} = F_n(x-x_\mathrm{dwarf},y-y_\mathrm{dwarf}) + F_\mathrm{ noise}
\end{equation}
along the $X$ axis (the equivalent equation holds for the $Y$ axis), where { $x_\mathrm{F}=-s_0(x-x_\mathrm{dwarf})$ are the FORS2 positions expressed in units of arc and measured relative to the approximate position $x_\mathrm{dwarf}$ of the target, $F_n(x-x_\mathrm{dwarf},y-y_\mathrm{dwarf})$}  is a sum of $M$ two-dimensional basic functions of the maximum power $n$, with free model coefficients,  which describe the transformation between the coordinate systems \citep{Lazorenko2009}, { that is between the local reference frame in some sky field and Gaia}. The term $F_\mathrm{ noise}$ models random noise that is not correlated across the field. The variance  of $F_\mathrm{ noise}$ is equal to the quadratic sum of $\sigma_\mathrm{ F}$ and the positional uncertainty in the external catalog.

\subsection{ FOV data set  of  FORS2 field stars}{\label{cat_raw_fov}}
A second result of these FORS2 observations are the astrometric  parameters of stars obtained also with $n=4$, but in a slightly different way. Specifically, each  star was reduced with its own circular reference area. Therefore the locations of the reference {frames} are not fixed and move in the CCD plane.  These individual areas cover the whole field of view of the CCD and therefore we will refer to it as FOV data set.  

The system of astrometric parameters in the FOV data set is not exactly homogeneous because the set of reference stars is different for every star, thus we are using multiple reference frames. In comparison to RF, the FOV data set data is larger and contains about 12\,000 stars,  because  it includes  stars outside the { central reference areas aligned to the program targets.}  These data were converted to the ICRF system with use of the USNO-B catalog \citep{USNO}  and are available in the CDS  \citep{2014yCat} as a deep catalog of positions, proper motions, and parallaxes of faint stars in 20 sky fields. Because that conversion was based on USNO-B, the absolute precision of the astrometry degraded to $\sim$0.2\arcsec, which is insufficient for the comparison with Gaia DR1. Therefore, to take advantage of the sub-mas precision of FORS2 differential astrometry, we instead used the original CCD positions and proper motions of FOV stars set by the reference frame fixed to the CCD pixels.

Both FORS2 data sets are based on exactly the same measured photocentres of star images and an identical reduction method but differ in the reference frame definition. Although this difference may appear subtle, the discussion in Sect.\,\ref{comp} demonstrates that the RF positions are in better agreement with Gaia. We present the analysis of both data sets because the FOV data set has a twice larger number of stars and therefore allows for a more robust comparison with DR1 positions.

\section{Data analysis}{\label{comp}}
We compared the positions of stars in Gaia DR1 that are in common with the RF and FOV data sets, which were converted from  their average epoch { 2011.5--2012.1 } to the Gaia DR1 epoch 2015.0 using the proper motions {measured with FORS2}. {These FORS2 positions refer to the barycentre of the Solar system, because the proper motions and parallaxes were accounted for and atmospheric chromaticity parameters were incorporated in the astrometric model.} The position uncertainties $\sigma_\mathrm{F}$ in 2015 were computed using the uncertainty and covariance matrix of the astrometric parameters of FORS2 stars, and includes input from all known types of errors. The average value of $\sigma_\mathrm{ F}$ over all sky fields is given in Table~\ref{table}. Because the proper motions of FORS2 catalogs were obtained with a short timebase { of about 2 years},  the conversion between epochs significantly degraded the precision of FORS2 positions. For example, the typical precision for bright stars  is roughly 0.1~mas for  positions and 0.07--0.15~mas/yr for proper motions. After conversion to 2015, the proper motion errors propagated for a  3.5 years difference increase the uncertainty of positions  four-fold, to about 0.4~mas.

\subsection{Error budget and unbiased RMS of the residuals}{\label{model}}
{We define the residuals, e.g.\ along the $X$ axis, between FORS2   $x_\mathrm{F}$ and DR1 positions  $x_\mathrm {abs}$ computed as a tangent projection to the CCD plane, as $\Delta=x_\mathrm{F}-x_\mathrm {abs}$ that indicates the left side of Eq.~(\ref{eq:abs}). The  expected variance of the residuals is modeled by} the sum $ {\sigma_\mathrm{ G}^2  + \sigma_\mathrm{ F}^2  }$ which includes the formal uncertainty $\sigma_\mathrm{ G}$ of DR1 and  the  uncertainty  $\sigma_\mathrm{ F}$ of FORS2 positions at 2015.  The  average value  of  $\sigma_\mathrm{ F}$   is 0.64~mas  and  small or comparable  to $\sigma_\mathrm{ G}$ for stars cross-identified with DR1. It is reasonable to assume that the measured variance  of ${\Delta}$ can deviate  from this model and a  more realistic error budget  is
\begin{equation}
\label{eq:seff}
  \sigma_{\Delta}^2 = {\sigma_\mathrm{ G}}^2  +  \sigma_\mathrm{ F}^2 + \nu^2,
\end{equation}
where we introduced an additional noise component $\nu$ to balance the discrepancy between the observed RMS of the residuals and the nominal  uncertainties in the catalogs. The $\nu$ component may be related to Gaia DR1 or FORS2 or both.

{The problem is that the residuals $\Delta$ cannot be measured  directly  because $x_\mathrm{F}$ and $x_\mathrm {abs}$ are related by the polynomial relation in Eq.\ (\ref{eq:abs}). While $x_\mathrm {abs}$ represents a flat coordinate system, the FORS2 positions $x_\mathrm{F}$ are affected by high-order geometric distortion. We therefore considered the corrected position residuals and the corrected residual RMS. The corrections are based on the fact  that the  unbiased  variance of the residuals ${\hat{\Delta}}_i=x_\mathrm{F}-x_\mathrm {abs}-F_n(x-x_\mathrm{dwarf},y-y_\mathrm{dwarf})$ computed  with the least squares fit of  $i=1\ldots N$ measurements { per axis for $N$ common stars} is   $\sum {\hat{\Delta}}^2_i w_i/(N-M)$, where $M$ is the number of fit parameters  and $w_i$ are the weights which depend on a factor (e.g. on $\sigma_\mathrm{ G}$)}. For a limited sub-sample of $N'<N$ residuals in a narrow range of $\sigma_\mathrm{ G}$ where the $w_i$ are approximately constant, the  unbiased estimate of the variance  is ${\langle {\hat{\Delta}}^2 \rangle N/(N-M)}$, where  the angle brackets denote an average taken over $N'$ measurements. Therefore the unbiased estimate of the residual RMS is $\mathrm{ RMS}=  \gamma {\langle {\hat{\Delta}}^2 \rangle }^{1/2}$  where  $  \gamma = \sqrt{N/(N-M)}$ is a factor that compensates for the decrease in the degrees of freedom. 

{ The original  residuals $\Delta$ are affected by the least squares fit and cannot be restored exactly. However, they can be approximated by the corrected residuals} $\Delta =  \gamma \hat{\Delta} = \gamma (x_\mathrm{F}-x_\mathrm {abs}-F_n)$ with  an unbiased variance.  In the following discussion we  deal  with the bias-corrected RMS and residuals $\Delta$ defined in this way.

\subsection{Individual residuals between DR1 and FORS2}{\label{residuals}}                       
Equatorial star positions of DR1 stars were converted  to Cartesian coordinates in the CCD plane {using a tangent-plane projection. The reference point of the projection corresponded to the position of target objects, i.e.\ to the geometric centres of the RF data sets. } Then we proceeded with  stars whose  uncertainty $\sigma_\mathrm{ G}$ is better than 20~mas,  cross-identified  DR1 and FORS2 stars, and applied  the polynomial model (Eq.\ \ref{eq:abs})  to transform between these two sets of positions. The data for FORS2 chip1 and chip2 were reduced separately  because the relative offset and orientation between the chips are known only approximately.  Cross-matching between catalogs consisted in the initial rough identification in a 1\arcsec\ window and three iteration cycles of fitting the residuals $\Delta$ by polynomial functions of degree $n=3$ and rejection of outliers over 60\,mas. With two concluding iterations, detailed below, we obtained the final identification and derived the residuals $\Delta$.

The main parameters of the fitting procedure are shown in Table\,\ref{table}. 
We present parameters for chip1 only because those for chip2 are similar, except for a smaller star number $N$ and polynomial degree $n$. For every field { and chip}, we had to choose the optimal polynomial degree $n$ with number $M$ of fit parameters.  Naturally, an increase of $M$ will lead to smaller residuals because there are more free parameters. Fitting polynomials with arbitrarily high degree is  undesirable and it is therefore necessary to identify the highest $M$ for a particular dataset.  Since our model is linear and the errors are reasonably well-behaved, we used the F-test of additional model parameters { which }yields the probability that the simpler model is true. This approach is described in detail by \cite{JWST-STScI-005492} who applied it to the comparison between Gaia DR1 and Hubble Space Telescope (HST) observations of the JWST calibration field. 

The weights in the  system Eq.\ (\ref{eq:abs}) of the transformation of FORS2 to DR1 were set to $  \sigma_ {\Delta}^{-2}$ initially computed with $\nu=0$, which therefore assumes that the uncertainties $\sigma_\mathrm{ G}$ and  $\sigma_\mathrm{ F}$ are good estimates. The stars were considered identified when  the residuals in positions were within $ 5\sigma_{\Delta}$.  Still, many stars were rejected as outliers because their residuals were slightly above the adopted 5-sigma limit.  The unusually high rate of outliers detected at this phase of identification, especially for stars with $\sigma_\mathrm{ G}<1$~mas, indicates that the distribution of the residuals deviates from the prediction (\ref{eq:seff}). To obtain a more stable  identification and to find a reasonable compromise between the number of identified stars and the RMS of the residuals, we therefore computed the final weights in most fields with different values of $\nu=[0, 0.7, 1]$\,mas.

\subsection{ RMS of the residuals }
\begin{figure}[h!]
\includegraphics[width=\linewidth]{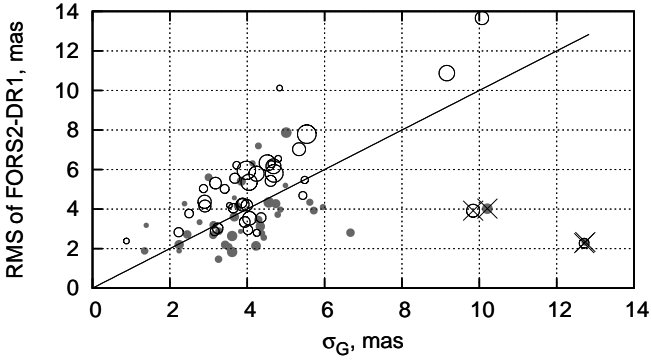}
\caption{ Average RMS of  the residuals $\Delta$  between DR1 and FORS2 data sets FOV (open  circles) and RF (gray circles) versus  { average value of} $\sigma_\mathrm{ G}$ for each sky field (field No 20 marked by crosses) and each chip. { The circle size is proportional to the number of stars, from 11 to 285}.}
\label{linear}
\end{figure}

{\label{rms}}   
The number of stars cross-identified with the RF and FOV data set is 1421 and 3705, respectively, and the residual scatter is smaller for the comparison of DR1 with RF. To illustrate the statistics of the residuals, we computed for every field and FORS2 chip the RMS of the residuals $\Delta$ and the (quadratic) average value  of $ \langle \sigma_\mathrm{ G}\rangle$, see Table \ref{table}. These values are visualised in Fig.\,\ref{linear} and show an approximately linear correlation at $2<\langle \sigma_\mathrm{ G}  \rangle<5$~mas where  $\sigma_\mathrm{ G}$ dominates over the other error components, including the FORS2 noise $\sigma_\mathrm{ F}$ that has an amplitude below 1~mas (Table \ref{table}).
The RMS  is systematically higher for the residuals between DR1 and FOV in comparison to the residuals between DR1 and RF data set, which is due to the difference of their construction as discussed in Sect.\ \ref{cat_raw_fov}. 
{ Note that field No 20 has unexpectedly small residuals at large $ \sigma_\mathrm{ G}$}.

\begin{figure}[htb]    
\begin{tabular}{@{}c@{}}
\resizebox{\hsize}{!}{\includegraphics*[width=\linewidth]{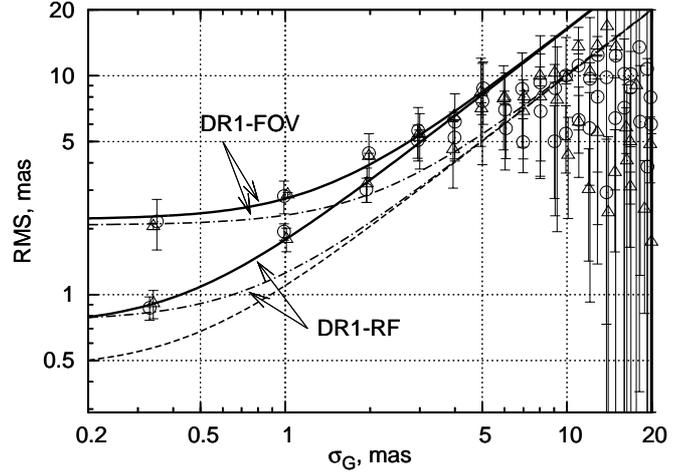}}\\
\end{tabular}
\caption{ RMS of the positional residuals $\Delta$ between DR1 and FOV  and between DR1 and RF   for chip1 (circles) and chip2 (triangles) in every 1~mas bin of $\sigma_\mathrm{G}$.  3-sigma error bars are drawn under the assumption of a normal error distribution. These data are compared with the model uncertainty $\sigma_{\Delta}$ computed with $\nu=0$ in Eq. (\ref {eq:seff}) (dashed lines) and with a value of $\nu$ that fits the RMS at $\sigma_\mathrm{ G}<0.5$~mas  (dashed-dotted lines).  Solid lines  show the fit function Eq. (\ref {eq:mod2}) .}           
\label{ch35}
\end{figure}

We investigated  how well the measured RMS matches the model precision $\sigma_{\Delta}$  along a full range of  $\sigma_\mathrm{ G}=0 \ldots 20$~mas. For this purpose the residuals FORS2$-$DR1 were binned in 1~mas intervals of $\sigma_\mathrm{ G}$ and quadratically averaged separately for every field and chip. The results presented in Fig.~\ref{ch35} demonstrate { significant deviations from} the model dependence Eq.~(\ref {eq:seff}) computed  with $\nu=0$ and shown by the dashed line, especially for $\sigma_\mathrm{ G}<0.5$~mas. In that region,  associated with brighter stars,   $\sigma_\mathrm{ G}$ and $\sigma_\mathrm{ F}$ are much smaller than the measured RMS values. This allowed us to derive a reliable estimate of $\nu$ by fitting the excess in RMS with the  expression Eq. (\ref {eq:seff}). In this way we obtained  $\nu=2.04 \pm 0.11$~mas  for the comparison of DR1 with FOV and   $\nu=0.66 \pm 0.08$~mas for the comparison of DR1 with RF.  The significantly smaller  excess $\nu$  in the latter case is a consequence of the more homogeneous astrometric system of the RF data set. The error component $\nu$ can be due to unaccounted excess noise in either DR1 or FORS2 astrometry, and the smaller value of $0.66 \pm 0.08$~mas can be put forward as {\it an upper limit of $\nu$ potentially related to DR1.} This estimate is strictly valid only for stars  with  G-band photometric magnitudes  of 16.0--17.5,  and average values of $\sigma_\mathrm{ G}\simeq0.31$~mas and $\sigma_\mathrm{ F}\simeq0.38$~mas.  The value $\nu=2.04$~mas  obtained from the comparison of DR1 with FOV is likely caused by noise related to the multiple reference frames used in the compilation of the FOV data set (Sect.\ \ref{cat_raw_fov}).

{ Since we determined the value of $\nu$, Eq.~(\ref {eq:seff}) is fully defined in the range of  $\sigma_\mathrm{ G}$ and for any star sample we can compute   $\sigma_\Delta$ which models  the expected value of the RMS.  The functional dependence of   $\sigma_\Delta$ on  $\sigma_\mathrm{ G}$ { is  shown by the dashed-dotted  lines in Fig.~\ref{ch35}. The measured RMS values now are well fit with the model at $\sigma_\mathrm{G}<$1~mas. However  at $\sigma_\mathrm{G}=$1--5~mas we note a systematic positive bias  of about a factor 1.5 which remains constant in the logarithmic scale of the plot.}}  Figure \ref{ch35} suggests that the model Eq.~(\ref {eq:seff}) is not adequate in the $\sigma_\mathrm{G}=1-10$~mas range  because here  we found $\mathrm{RMS}>\sigma_{\Delta}$. In this interval  $\sigma_\mathrm{G}$ is much larger than the other noise components in Eq.~(\ref {eq:seff}),  including $\nu$ and $\sigma_\mathrm{ F}$ which typically are 0.5--0.7~mas. Therefore, the discrepancy with the observations is likely due to an underestimated variance $\sigma_\mathrm{G}^2$. We laid out the improved error budget 
\begin{equation}
\label{eq:mod2}
  \sigma_{\Delta}^2 = {A^2\sigma_\mathrm{ G}}^2  +  \sigma_\mathrm{ F}^2 + \nu^2 
\end{equation}
with  an additional parameter $A$ that modulates the DR1 uncertainties. $A\!>\!1$ and $A\!<\!1$ indicate that $\sigma_\mathrm{G}$ is under- and over-estimated, respectively.  We fitted the complete sample of all residuals with this model, where we ingested both data sets of FORS2 and used three fit parameters: $\nu_\mathrm{ FOV}$  for the residuals 'DR1'-'FOV',  $\nu_\mathrm{ RF}$ for the residuals 'DR1'-'RF',  and $A$ as a common parameter. The data were fit  in the  $\sigma_\mathrm{G}$ range 0 -- 5.5~mas and produced the model parameters  $\nu_\mathrm{ FOV}=2.13 \pm 0.09  $~mas,   $\nu_\mathrm{ RF}= 0.44 \pm 0.13$~mas,  close to the previous estimate, and  $A=  1.63  \pm  0.05$. The fit approximates the observed data well for $\sigma_\mathrm{G}<5$\,mas. At larger $\sigma_\mathrm{G}$ in the interval of 7--20~mas, the measured RMS reaches a ceiling of about 5--8~mas, therefore Eq.~(\ref{eq:mod2}) is not applicable there.

{ The solution of Eq.~(\ref{eq:mod2}) differs between data sets in the range of  $\sigma_\mathrm{G}<1$\,mas because there $\sigma_\mathrm{F}$ is comparable to $\sigma_\mathrm{G}$. We fit the observed RMS in this region separately for both FORS2 data sets but obtained similar solutions for $A$. Therefore, we fit the combined (FOV and RF) data set with  Eq.~(\ref{eq:mod2}) and obtained  $A=  1.83  \pm  0.07$, similar to the above result obtained in a wider range of $\sigma_\mathrm{G}$. }It means that  Eq.~(\ref{eq:mod2}) is applicable at least within a range of $0.5<\sigma_\mathrm{G}<5$~mas and our results suggest that in this region the uncertainties of the DR1 secondary data set are underestimated. 

\begin{figure*}[tbh]
\begin{tabular}{@{}c@{}c@{}c@{}}
{\includegraphics[height=4 cm, width=6.3cm]{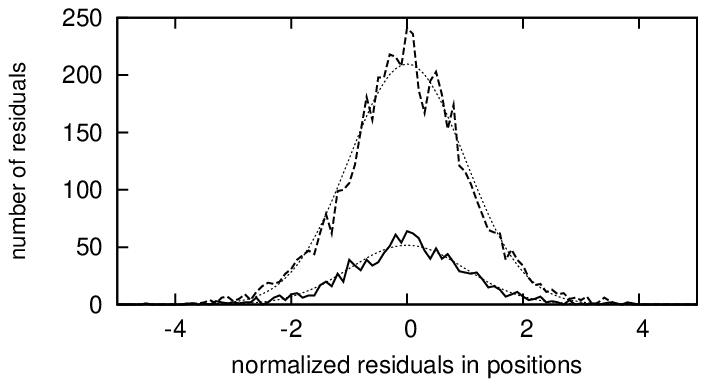}}&
{\includegraphics[height=4 cm, width=6cm]{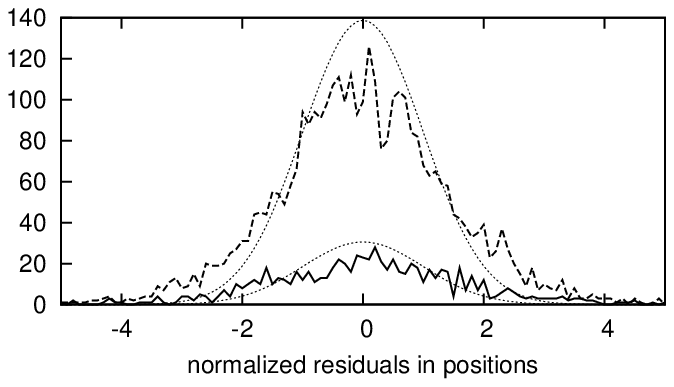}}&   
{\includegraphics[height=4 cm, width=6cm]{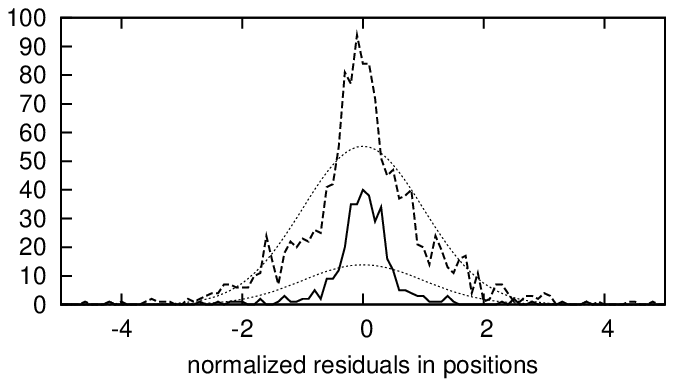}}
\end{tabular}
\caption{Distribution of the normalised position residuals $\Delta$ between  DR1 and RF (solid curves) and between DR1 and FOV (dashed curves) for stars with $1<\sigma_\mathrm{G}<5$~mas. \textit{Left panel}:  The normalisation is based on  $\sigma_{\Delta}$ computed with $A=1.63$;  \textit{Middle panel:}  The same with $A=1$;   \textit{Right panel}: The same for stars with $\sigma_\mathrm{G}>7$~mas   assuming that  $A=1$. The corresponding Gaussian distributions are shown by dotted lines.}
\label{hist}
\end{figure*}

We {studied whether there is} a difference in the residual RMS for stars with large and small \textit{astrometric excess noise} $\varepsilon$, which indicates  the quality of the astrometric fit for every star  in DR1  \citep{DR1}. The stars were divided in two groups  with `large` and `small` $\varepsilon$ as defined by a threshold value.  Using different thresholds from 0.5 to 2~mas, we found no clear difference in the RMS value for the subsets of stars, partly due to the small number of stars with large $\varepsilon$. 

{ We found that Eq.(\ref{eq:mod2}) fits well the measured RMS in every field, except for} field No 20 with an RMS significantly below $\sigma_{\Delta}$ and 3--5 times smaller than $\sigma_\mathrm{G}$.  This field is unusual in terms of the stellar brightness and $\sigma_\mathrm{ G}$ distributions. Whereas in most fields only $\sim$10\% of stars have $\sigma_\mathrm{ G}$ values of 10 -- 20~mas, about half of the stars in field No 20 have such large uncertainties. This is despite those stars are relatively bright ($G\simeq 17-19$) and should therefore have $\sigma_\mathrm{ G} \simeq0.5-2$~mas as typical in the other fields.  The reason for this peculiarity may be  that Gaia observed this field less often, {as reported by the average number of good CCD observations in DR1 (catalog entry \emph{astrometric\_n\_good\_obs\_al}), which is the smaller than for the other fields}. {Our findings therefore} may demonstrate variations of the DR1 uncertainties in different areas in the sky.

\subsection{Distribution of the residuals }{\label{distr}}   
The position residuals contain information on the distribution of DR1 errors, which can deviate from the normal law.  Figure \ref{hist} shows  the observed distribution of the residuals between   FORS2 data sets and DR1  normalized to $\sigma_{\Delta}$ computed with the derived fit parameters. Because  the uncertainty $\sigma_{\Delta}$ is correctly modeled by  Eq.~(\ref{eq:mod2}) only within a limited range of $\sigma_\mathrm{G}$, we generated the histograms for stars in a conservative $1<\sigma_\mathrm{G}<5$~mas range. We rejected residuals with large correction factors $\gamma>1.3$,   a situation usually related to sky fields with insufficient  number of identified stars $N$. The derived distributions follow a Gaussian distribution with no excess in the  wings.

To demonstrate how the distribution shape depends on the value of $A$, we computed the uncertainties  $\sigma_{\Delta}$ with $A=1$ (that is using the original uncertainties of DR1) and obtained the distributions in the middle panel of Fig.~\ref{hist} with the best-fit estimates of $\nu_\mathrm{ FOV}$ and $\nu_\mathrm{ RF}$. The difference relative to the left  panel  demonstrates the effect of the bias in the $\sigma_{\Delta}$. Finally, we obtained the residual distribution for stars with $\sigma_{G}>7$~mas (right panel). The strong  concentration at zero indicates an overestimation of $\sigma_{G}$ by at least a factor of two, as is also seen in Fig.~\ref{ch35}.

\subsection{Discussion}
It is unlikely that the entirety of these results can be caused by errors in the FORS2 astrometry. For instance,  excess noise $\nu$ can also originate if $\sigma_\mathrm{ G}$ is correct but instead our value of $\sigma_\mathrm{F}$ is underestimated. For stars with $\sigma_\mathrm{ G}<1$~mas the average value of  $\sigma_\mathrm{F}$ is 0.41~mas. Increasing this value quadratically by $\nu=0.44$~mas yields a new estimate of  $\sigma_\mathrm{F}= 0.73$~mas   (78\% over of its nominal value), which removes the discrepancy between the measured and model variance of the residuals FORS2$-$DR1, while keeping $\sigma_\mathrm{ G}$ untouched in this range.  In this scenario, because $\sigma_\mathrm{ F}$ is dominated by the uncertainty in proper motion, the 78\% excess in our errors would refer almost entirely to  FORS2  proper motions. However, applying the same argument in the range of $\sigma_\mathrm{ G}>1$~mas, the excess noise $\nu$ cannot be due to underestimated FORS2 errors because { that} would lead to unrealistic manifold corrections to $\sigma_\mathrm{F}$.

Potential sources of unaccounted noise in FORS2 positions are systematic errors in the measured photocentres of stars caused by imperfect modeling of the point spread function or from unmodelled blended light from nearby stars. We investigated other potential sources of the excess noise by looking for correlations between the residuals and brightness, position on the CCD, proper motion, chromaticity parameters of individual stars. We did not find any significant correlation.

\section{Update of the FORS2 pixel scale}{\label{parall}}
The conversion of relative to absolute positions on the basis of Gaia DR1 allows us to derive the coefficients of the function $F_n(x-x_\mathrm{dwarf},y-y_\mathrm{dwarf})$, which {determine} the optical distortions of the FORS2 camera, the perpendicularity of the CCD axes, the pixel scale and  its variation across the CCD, and the differences of scales along $X$ and $Y$ directions. It is convenient to present Eq.\,(\ref{eq:abs}) in the conventional form  $s_0 x' - x_\mathrm{G}  = ax'+by'+{x'}_0 + \ldots $ and $s_0 y' - y_\mathrm{G}  = cx'+dy'+{y'}_0 + \ldots  $ where $x'=-(x-x_\mathrm{dwarf})$,  $y'=y-y_\mathrm{dwarf}$ and the high-order terms of the function $F_n$ were omitted. Here we define the scale $s$ and the coefficients of geometric distortion as
\begin{equation}
\begin{array}{lcr}
 s  = s_0-(a+d)/2, \quad
 s_x -s_y = (d - a), \quad 
 skew = (b + c) 
\end{array}  
\end{equation}
where   $s_x -s_y$  characterises  the scale difference between the two axes and the $skew$ term gives information on the non-perpendicularity of those axes. Above expressions are applicable at the location of  the target, where  $x'=y'=0$. { In general, $s(x,y)$ and the other distortion coefficients are  functions of field location and can be computed from the partial derivatives of $F_n$ including the higher order terms \citep{JWST-STScI-005492}. }

\begin{figure}[htb]    
\begin{tabular}{@{}c@{}}
\resizebox{\hsize}{!}{\includegraphics*[width=\linewidth]{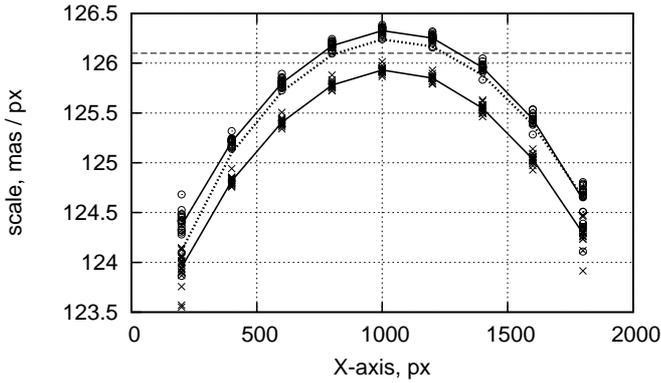}}\\
\end{tabular}
\caption{The change of FORS2 pixel scale $s$ in chip1 along the $X$ axis for each sky field at $Y=100$~px (open circles) and at $Y=500$~px (crosses). The average scales (solid curves), the  scale obtained by \citet{PALTA2} (horizontal dashed line),  { and the scale $s$  computed with use of distortion coefficients in chip 2 at $Y=0$ (dotted curve) are shown.}}           
\label{scale2}
\end{figure}

\begin{table*}[tbh]
\caption [] {Pixel scales, the skew and rotation parameters, and updated ultracool dwarf parallaxes  $\varpi$}
{\footnotesize
\centering
\begin{tabular}{@{}rcccccccc@{}}
\hline
\hline
Nr &DENIS-P            &$s_0   $         & $s_\mathrm{FOV}$  & $s_\mathrm{RF}$       &  $\varpi_0$  &     $\varpi$        &   $skew$       &  $rotation$    \rule{0pt}{11pt}   \\
   &                   & (mas/px)        &  (mas/px)           &  (mas/px)           &  (mas)    & (mas)                 &   (mas/px)      &    (mas/px)    \\
\hline                                                                                                                                                                                                            
1 & J0615493-010041    & 126.17$\pm$0.13 & 126.279$\pm$0.006   & 126.295$\pm$0.004   &    45.700 & 45.745   $\pm$0.112  & 0.034$\pm$  0.006 &   0.166 $\pm$  0.013   \rule{0pt}{11pt}   \rule{0pt}{11pt}  \\
2 & J0630014-184014    & 126.20$\pm$0.11 & 126.333$\pm$0.013   & 126.329$\pm$0.015   &    51.719 & 51.772   $\pm$0.099  & 0.002$\pm$  0.024 &   0.215 $\pm$  0.017   \\
3 & J0644143-284141    & 126.10$\pm$0.14 & 126.330$\pm$0.005   & 126.331$\pm$0.003   &    25.094 & 25.140   $\pm$0.094  &-0.002$\pm$  0.004 &   0.193 $\pm$  0.008   \\
4 & J0652197-253450    & 126.08$\pm$0.12 & 126.265$\pm$0.019   & 126.307$\pm$0.006   &    62.023 & 62.135   $\pm$0.070  & 0.008$\pm$  0.010 &   0.206 $\pm$  0.021   \\
5 & J0716478-063037    & 126.45$\pm$0.14 & 126.355$\pm$0.008   & 126.324$\pm$0.005   &    40.918 & 40.877   $\pm$0.144  & 0.053$\pm$  0.009 &   0.201 $\pm$  0.010   \\
6 & J0751164-253043    & 126.24$\pm$0.12 & 126.325$\pm$0.013   & 126.327$\pm$0.004   &    56.304 & 56.343   $\pm$0.085  &-0.030$\pm$  0.006 &   0.213 $\pm$  0.015   \\
7 & J0805110-315811    & 126.07$\pm$0.11 & 126.359$\pm$0.013   & 126.337$\pm$0.008   &    42.428 & 42.518   $\pm$0.083  & 0.015$\pm$  0.012 &   0.267 $\pm$  0.015   \\
8 & J0812316-244442    & 126.01$\pm$0.09 & 126.332$\pm$0.008   & 126.322$\pm$0.003   &    47.282 & 47.399   $\pm$0.094  & 0.048$\pm$  0.005 &   0.188 $\pm$  0.014   \\
9 & J0823031-491201    & 126.26$\pm$0.13 & 126.363$\pm$0.018   & 126.331$\pm$0.005   &    48.160 & 48.187   $\pm$0.190  &-0.032$\pm$  0.006 &   0.203 $\pm$  0.021   \\
10&  J0828343-130919   & 126.18$\pm$0.09 & 126.353$\pm$0.008   & 126.355$\pm$0.006   &    85.838 & 85.957   $\pm$0.148  & 0.012$\pm$  0.017 &   0.170 $\pm$  0.018   \\ 
11&  J1048278-525418   & 126.01$\pm$0.08 & 126.339$\pm$0.008   & 126.345$\pm$0.003   &    36.212 & 36.308   $\pm$0.077  & 0.101$\pm$  0.005 &   0.268 $\pm$  0.013   \\
12&  J1157480-484442   & 126.23$\pm$0.11 & 126.344$\pm$0.014   & 126.354$\pm$0.005   &    34.633 & 34.667   $\pm$0.082  & 0.009$\pm$  0.012 &   0.223 $\pm$  0.021   \\
13&  J1159274-524718   & 126.32$\pm$0.06 & 126.352$\pm$0.010   & 126.322$\pm$0.003   &   105.538 &105.540   $\pm$0.120  &-0.012$\pm$  0.007 &   0.202 $\pm$  0.014   \\
14&  J1253108-570924   & 126.00$\pm$0.10 & 126.274$\pm$0.026   & 126.207$\pm$0.034   &    60.064 & 60.163   $\pm$0.054  &-0.093$\pm$  0.038 &   0.128 $\pm$  0.026   \\
15&  J1520022-442242   & 126.08$\pm$0.09 & 126.321$\pm$0.017   & 126.316$\pm$0.009   &    53.995 & 54.096   $\pm$0.109  & 0.057$\pm$  0.018 &   0.239 $\pm$  0.023   \\
16&  J1705474-544151   & 126.12$\pm$0.07 & 126.358$\pm$0.011   & 126.293$\pm$0.011   &    37.549 & 37.601   $\pm$0.087  & 0.041$\pm$  0.017 &   0.205 $\pm$  0.018   \\
17&  J1733423-165449   & 126.46$\pm$0.09 & 126.355$\pm$0.013   & 126.333$\pm$0.010   &    55.272 & 55.216   $\pm$0.073  & 0.037$\pm$  0.016 &   0.174 $\pm$  0.016   \\
18&  J1745346-164053   & 126.09$\pm$0.26 & 126.299$\pm$0.010   & 126.294$\pm$0.006   &    50.871 & 50.953   $\pm$0.096  & 0.034$\pm$  0.010 &   0.201 $\pm$  0.018   \\
19&  J1756296-451822   & 125.97$\pm$0.12 & 126.244$\pm$0.025   & 126.318$\pm$0.011   &    43.577 & 43.697   $\pm$0.064  & 0.194$\pm$  0.018 &   0.247 $\pm$  0.040   \\
20&  J1756561-480509   & 126.30$\pm$0.13 & 126.292$\pm$0.018   & 126.266$\pm$0.008   &    47.039 & 47.026   $\pm$0.058  &-0.018$\pm$  0.012 &   0.184 $\pm$  0.019   \\
\hline                                                                                                                                    
  &   median           & 126.14$\pm$0.11 & 126.336$\pm$0.010   & 126.323$\pm$0.008   &           &                      &      $\sim 0$      &   0.202$\pm$0.016      \\
\hline                                                                                                                                    
 \multicolumn{2}{c}{  Luhman 16  }& 126.10$\pm$0.11 & \multicolumn{2}{c}{126.329$\pm$0.010   } &   500.51  &501.419    $\pm$0.11  &        &\\       
\hline                 
\end{tabular}   
\label{table2}          
}                      
\end{table*}           
The values of $s$ for each sky field at the position of the target $x=x_\mathrm{dwarf}$, $y=y_\mathrm{dwarf}$ are given in Table\,\ref{table2} both for RF and FOV data sets. For comparison, the table contains scales $s_0$ which were previously derived on the basis of the USNO-B catalogue \citep{PALTA2}. Those were smaller than $s_\mathrm{FOV}$ and $s_\mathrm{RF}$. This underestimation of the pixel scale was { caused} by the small degree of the polynomial transformation $F_n$ that we used in \citep{PALTA2} that was fixed to $n=3$ for all fields, but did not carry the $x^2 y$ and $x y^2$ terms because of the insufficient precision of USNO-B. That function $F_n$ allowed for the transformation to ICRF with  a precision of   40--70~mas, however, the absence of the $x^2 y$, $ xy^2$, and higher-order terms prevented it to resolve the field distortion. This is evident in Fig.~\ref{scale2} which shows how the scale $s(x,y)$ changes along the $X$ axis of chip1 at $Y=100$~px, close to the chip gap. The figure corresponds to the FOV data set, but it is nearly the same for the RF set. The function $s(x,y)$  is symmetric and reaches  a maximum value of 126.33~mas at the chip center (column $X=1000$~px). The use of a simpler transformation $F_n$  broadened the peak of $s(x,y)$  and yielded the lower value of 126.1~mas reported in \citep{PALTA2}.

The function $F_n$ represents the difference between the local reference frame and Gaia, but not directly between the star positions in the CCD system and Gaia. This induces additional distortions in the differences  $x-x_\mathrm{dwarf}$ and $y-y_\mathrm{dwarf}$ and is the reason why the individual values of $s$ are scattered by over 3-sigma, see Table \ref{table2}. In particular, the scatter of points in Fig.~\ref{scale2} is due to this effect. Fig. \ref{scale2} shows that the scale changes across the chip and in the outer regions of $Y=500$~px it is about 0.5 mas/px lower compared to at $Y=100$~px. This confirms that the conversion from pixel to angular units should be done with a scale evaluated at the position $x=x_\mathrm{dwarf}$, $y=y_\mathrm{dwarf}$ of the target. In our observations $x_\mathrm{dwarf}$ and $y_\mathrm{dwarf}$ varied by $\sim100$ pixels for different targets, thus adding additional scatter in the measured scale (Table\,\ref{table2}) because that is a function $s(x,y)$. 

{ The pixel scales in chip 1 and chip 2 of FORS2 cannot be directly compared  because  they are  modulated by the optical distortion. To investigate possible scale differences between chips, we used the coefficients of $F_n$ of chip 2 and computed $s(x,y)$ at the dividing gap between the chips, where $F_n$ is still applicable.  These scales  correspond to Y=0 in chip 1 (dotted { curve}   in  Fig.~\ref{scale2}) and are in a good agreement with scales at the nearby location defined by $Y=100$~px. Attributing the common curved shape to the optical distortion, and the offset to the separation in $Y$, we find no scale discontinuity between the two chips.}

 The $skew$ parameter given in Table\,\ref{table2} represents the axes non-perpendicularity is  small (0.01--0.02 mas/px) and consistent with zero. However, it formally varies by over 3-sigma between the sky fields due to the skew of the local reference frames. { The difference $s_x -s_y $} shows a similar random scatter with an average close to zero. This disagrees with our former estimate of $-0.37\pm 0.06$~mas/px \citep{PALTA2}, possibly for the reason similar to the discussed difference of the pixel scale.

{ The FITS headers of FORS2 indicate a relative rotation of 0.083\degr between the two CCD chips. We verified this by evaluating the coefficients of the function $F_n$. For chip 1, we find the inclination  $\theta_1=(b-c)/2$ between the ICRF and the axes of the FOV reference frames at the position of the target, which  approximately corresponds to the inclination between the ICRF and the the CCD axes.  We used the function $F_n$ for chip 2 to determine the inclination $\theta_2$ about 200~px below the target. Instead of $b$ and $c$ we used their exact local values computed with the partial derivatives of $F_n$. The difference $\theta_1 - \theta_2$ between the inclination angles is the rotation between the chips, which is given in Table\,\ref{table2} for each field. Its median value is $(0.202 \pm 0.016)$~mas/px, thus the $Y$ axis of chip 1 is rotated clock-wise relative to that of chip 2. This corresponds to a rotation of $(0.092 \pm 0.009)\degr$ and agrees with the value given in the FITS headers.}

\section{Updated parallaxes of ultracool dwarfs}{\label{updated}}
With the now better-determined pixel scales we corrected the (absolute) parallaxes $\varpi_0$ of 20 ultracool dwarfs reported in \cite{Palta1} that were computed with scales $s_0$ given in Table\,\ref{table2} by applying a multiplicative factor $\beta= s/s_0$. We used scales obtained in individual sky fields to take into account  variations related to the local reference frames. We { used  $s= s_\mathrm{RF}$  because }of the better internal precision, but the factors $\beta$ computed { with $s =s_\mathrm{FOV}$ are} nearly the same. The applied $\beta$ factors vary from 0.9990 to 1.0028, thus the corrections are small and change the parallaxes $\varpi$ by about 0.1~mas, comparable to the formal parallax uncertainties.

The effect is more significant for the binary brown dwarf WISE J104915.57$-$531906.1 (LUH16, \citealt{Luhman}). By analysing the FORS2 images obtained by \citet{Boffin} we derived the relative (500.23\arcsec) and absolute parallax $\varpi=500.51 \pm 0.11$~mas of this system located $\sim$2 pc from the Sun \citep{2015MNRAS}. Using HST observations in 2014--2016, a pixel scale determined on the basis of Gaia DR1, {and the parallax correction of \citep{2015MNRAS}}, \citet{Bedin} derived the relative and absolute parallax of LUH16 of $501.118 $~mas and  $501.398 \pm 0.093$~mas, respectively, significantly larger than the values of \citet{2015MNRAS}.  Applying the correction factor $\beta$ computed with $s_0=126.1$~mas from  \cite{2015MNRAS} and $s=126.329 \pm 0.010$~mas, which is the median of the complete set of $s_\mathrm{FOV}$ and $s_\mathrm{RF}$, we obtained the updated relative 501.139~mas and absolute $\varpi=501.419 \pm 0.11$~mas FORS2 parallax of LUH16, in agreement with \citet{Bedin}.

\section{Conclusion}{\label{c_d}}
We compared the predicted and measured residual RMS in position between the Gaia DR1 and FORS2 astrometric data sets of 20 astrometric fields. The relationships between these two datasets are non-trivial and require the introduction of an auxiliary term $\nu$ to take into account the excess in the residual scatter. In addition, the behaviour depends on the Gaia position uncertainty $\sigma_\mathrm{ G}$ and is different in the ranges smaller than 5~mas and larger than 7~mas. Our study is sensitive to the random component of position errors that are uncorrelated on spatial scales of about 2--4\arcmin\ because of the limitations of the FORS2 differential astrometry.  Our results apply to the faint end of the Gaia DR1 content, i.e.\ to stars with $G>16$.

In the range of $\sigma_\mathrm{ G}=0.5-5$\,mas, our results suggest that the actual value of the DR1 uncertainty  is underestimated by 63\% for 80\% of stars typically fainter than $G=17$. This conclusion agrees with the finding of \citet[Appendix A,][]{Mignard2016} on the uncertainty of the Gaia DR1 secondary data set.  In contrast, we find that for $\sigma_\mathrm{ G}>7$~mas, i.e.\ mostly $G>20$ stars, the actual DR1 uncertainties are overestimated by a factor of two.

The excess noise $\nu$ in our model was detected primarily in the residual RMS of $G=16-18$ stars with $\sigma_\mathrm{ G}<0.5$~mas. This noise can originate from Gaia DR1 and/or from FORS2. We cannot {pinpoint} the likely source, but we determined that the component related to Gaia DR1  cannot exceed  $0.44 \pm 0.13$~mas, which corresponds to $\nu_\mathrm{ RF}$ for the RF data set.  {We expect that the situation will be clarified with the second Gaia data release that will provide us with astrometry of even higher quality.}

Finally, the availability of the Gaia astrometry allowed us to more precisely calibrate the geometric distortions of our extensive FORS2 data sets, to derive a better pixel scale, and consequently update the parallaxes of the 20 ultracool dwarfs in our planet search program and of the LUH16 binary. Whereas for most of our targets the correction is smaller than 0.1~mas, the updated FORS2 parallax of  LUH16 system is larger by about 1~mas. { This demonstrates the value of the accurate, optically faint, and dense reference frame that Gaia provides for high-precision ground-based differential astrometry.}

\begin{acknowledgements}
This work has made use of data from the ESA space mission \emph{Gaia} (\url{http://www.cosmos.esa.int/gaia}), processed by the \emph{Gaia} Data Processing and Analysis Consortium (DPAC, \url{http://www.cosmos.esa.int/web/gaia/dpac/consortium}). Funding for the DPAC has been provided by national institutions, in particular the institutions participating in the \emph{Gaia} Multilateral Agreement.
\end{acknowledgements}

\bibliographystyle{aa}
\bibliography{pgaia}

\end{document}